\documentclass{lmcs}

\AtBeginDocument{%
  }

\begin{document}

\title[CITYPULSE: Real-Time Traffic Data Analytics]{CITYPULSE: Real-Time Traffic Data Analytics and Congestion Prediction}

\author[Idriss D. Teledjieu]{Idriss Djiofack Teledjieu\lmcsorcid{0009-0004-9142-9036}}[a]
\author[Irzum Shafique]{Irzum Shafique\lmcsorcid{0009-0003-8061-1839}}[b]

\address[a]{University of Colorado Boulder, USA}
\email{idriss.djiofackteledjieu@colorado.edu}

\address[b]{Xidian University, China}
\email{irzum@stu.xidian.edu.cn}



\begin{abstract}
\textit{CityPulse} is a proof-of-concept big data pipeline designed to enable real-time urban mobility analytics using scalable, containerized components—without reliance on physical sensor infrastructure. The system simulates the ingestion of 11 million traffic-related records representing urban phenomena such as vehicle congestion, GPS coordinates, and weather conditions. Data is ingested through a Dockerized Apache Kafka cluster, coordinated by ZooKeeper, and processed in real time using Apache Spark Structured Streaming.

To ensure robustness under load, the architecture introduces a temporary data storage layer that buffers Spark output before committing it to a centralized data warehouse. This design improves write efficiency, fault tolerance, and enables batch processing of intermediate results. The refined data feeds into a lightweight machine learning module and is served through a Flask backend with a React-based frontend for visualization and interaction.

Stress testing shows that the system maintains over 300K records/min throughput with only a 10\% increase in latency under full load conditions. With its modular Docker-based deployment, \textit{CityPulse} offers a cost-effective and reproducible analytics solution for traffic congestion monitoring in resource-constrained environments, particularly in developing regions like Cameroon.
\end{abstract}


\maketitle

\begin{figure}[htbp]
  \includegraphics[width=\textwidth]{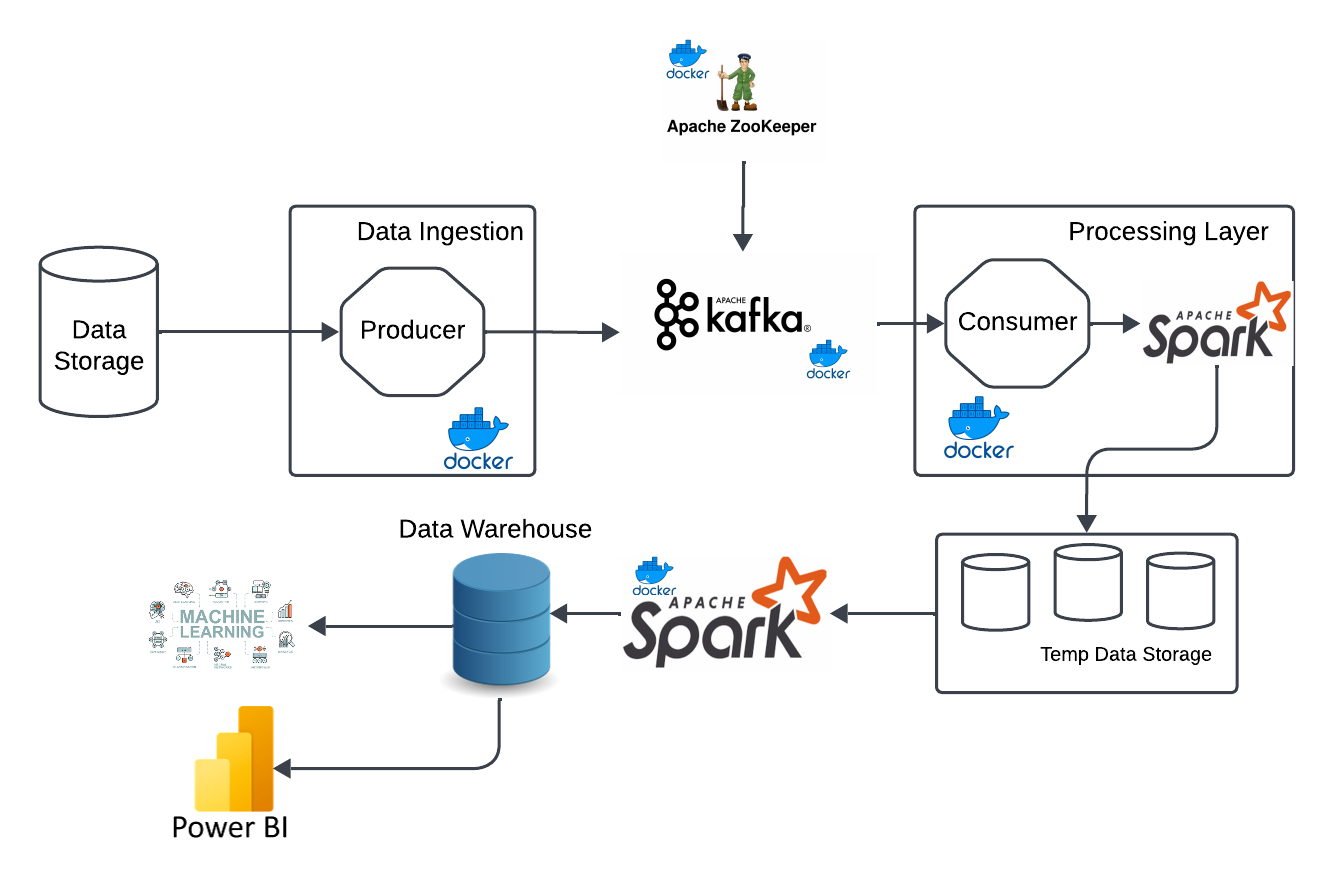}
  \caption{CityPulse: System Architecture}
  \label{fig:teaser}
\end{figure}

\section{Introduction}

Urban centers across the globe face growing challenges related to traffic congestion, inefficient mobility systems, and public safety—issues that are especially acute in developing regions such as Cameroon. Traditional traffic monitoring infrastructure typically depends on extensive, high-cost sensor networks~\cite{kheder2024real}, which are often impractical in these settings due to budgetary and infrastructural constraints.

To address this gap, \textit{CityPulse} presents a novel proof-of-concept solution that leverages a scalable, real-time data pipeline built entirely from open-source and containerized technologies. Rather than relying on physical sensors, the system generates synthetic data streams emulating urban traffic signals, such as vehicle telemetry, GPS coordinates, and weather patterns. This synthetic approach enables fast, low-cost experimentation and iteration.

\textit{CityPulse} employs Apache Kafka (for ingestion), ZooKeeper (for coordination), and Apache Spark (for streaming and transformation), all orchestrated via Docker. The system uses a temporary storage layer to decouple processing from warehousing, ensuring scalability and write efficiency. Final data is stored in a central warehouse, processed by a lightweight machine learning component, and exposed through a Flask API with a React frontend.

This architecture demonstrates how real-time traffic insights can be delivered efficiently, even in constrained environments. By eliminating the dependency on costly infrastructure, \textit{CityPulse} provides a practical and replicable model for smart city analytics in the Global South.

\section{Related work}
The increasing need for scalable and cost-effective solutions for traffic monitoring in urban areas has spurred research to use real-time data from various sources. Notable studies have demonstrated different approaches to leveraging these data for traffic management and event detection systems.

\begin{itemize}
    \item \textbf{Integration of IoT and Big Data for Traffic Monitoring}: \textit{Shi et al. (2015)} discussed the implementation of a Big Data-driven approach to real-time traffic and safety monitoring on urban expressways using Microwave Vehicle Detection Systems (MVDS). The system processed data at high temporal granularity to assess congestion and predict crash risks, showcasing the potential of using dense sensor deployment for comprehensive traffic analysis. \cite{shi2015big}

    \item \textbf{Use of Social Media for Event Detection}: \textit{D’Andrea et al. (2015)} utilized Twitter streams to detect real-time traffic events, demonstrating that social media could be effective in identifying traffic congestion and accidents before they are reported by traditional news outlets. This approach highlights the potential of social networks as a supplementary real-time information source for traffic management systems. \cite{d2015real}

    \item \textbf{Employment of Deep Learning Techniques in Traffic Systems}: \textit{Kheder and Mohammed (2024)} developed an IoT-aided robotics model that incorporates deep learning techniques for real-time traffic monitoring. Their system uses modified LeNet-5 and Inception-V3 models for traffic sign and light recognition, illustrating the effectiveness of combining IoT with advanced image processing techniques for dynamic traffic management. \cite{kheder2024real}

    \item \textbf{Real-Time Traffic Data Utilization}: \textit{Wang et al. (2024)} proposed the Traffic Performance GPT (TP-GPT), an intelligent chatbot that integrates large language models with traffic databases to provide real-time traffic management solutions. This system underscores the utility of real-time data and advanced analytics in enhancing traffic decision-making processes. \cite{wang2024traffic}

    \item \textbf{Cloud-Based Solutions for Traffic Data}: The concept of vehicular cloud computing (VCC) was explored by researchers as a method to utilize cloud technologies for providing scalable traffic management solutions. VCC uses the cloud to offer low-cost computing services to drivers, aiming to improve overall traffic flow and safety.
\end{itemize}

Recent studies highlight a growing shift toward integrating advanced technologies—such as IoT \cite{huang2019real}, big data analytics, cloud-native infrastructure, and artificial intelligence—into traditional traffic monitoring systems. This convergence enhances the ability of traffic management frameworks to cope with the increasing complexity and scale of urban mobility. Building on these developments, the proposed \textit{CityPulse} project implements a real-time data pipeline for smart city analytics using fully containerized, open-source tools. By combining synthetic data generation, real-time stream processing, and efficient storage and visualization mechanisms, the system offers a cost-effective alternative for cities in developing regions like Cameroon. This design not only addresses the infrastructural limitations common in such contexts but also positions them to adopt next-generation traffic intelligence solutions through scalable and replicable technology.

\section{Our Proposed Work}

\textit{CityPulse} is a real-time traffic analytics and congestion prediction framework designed to simulate and process large-scale urban mobility data using open-source technologies. The system ingests 11 million records of synthetic traffic data and classifies congestion levels (\textit{Low}, \textit{Medium}, \textit{High}) using a scalable, fully containerized pipeline. It is built to demonstrate how intelligent mobility insights can be delivered efficiently, especially in infrastructure-constrained settings such as Cameroon.

\subsection{System Architecture and Components}

The pipeline consists of the following core components, each deployed via Docker:

\begin{itemize}
    \item \textbf{Kafka Producer}: Reads raw traffic data (CSV), converts it to JSON, and streams it to the \texttt{raw-traffic-data} Kafka topic in batches of 500. It includes retry logic, message flushing, and Snappy compression to optimize throughput and reliability.
    
    \item \textbf{Kafka Consumer with Spark}: Buffers incoming Kafka messages in batches of 500 and applies data cleaning, feature engineering, and clustering using KMeans. Temporary results are stored locally before being written to the main data warehouse.
    
    \item \textbf{Apache Spark}: Performs structured streaming and transformation of raw data. Spark also supports model input preparation and batch scoring of congestion levels.
    
    \item \textbf{Machine Learning Module}: Trains a Random Forest classifier using key engineered features such as \texttt{v\_Vel}, \texttt{v\_Acc}, \texttt{Space\_Headway}, and \texttt{Time\_Headway}, with congestion labels generated via unsupervised KMeans clustering.
    
    \item \textbf{Flask + React Interface}: The backend exposes ML predictions via Flask, while the frontend built in React displays road congestion levels with simulated city and road names.
\end{itemize}

\subsection{Data Pipeline Functionality}

\begin{itemize}
    \item Dataset preprocessing includes:
    \begin{itemize}
        \item Filling missing values in velocity and acceleration with 0.0
        \item Replacing missing IDs (e.g., Lane\_ID, Section\_ID) with -1
        \item Defaulting gaps (Space\_Headway, Time\_Headway) to 0.0
    \end{itemize}
    
    \item Kafka producer sends data in \textbf{batches of 500}, enabling better control over flow and resource utilization.
    
    \item Kafka consumer performs real-time data transformation and KMeans clustering for unsupervised congestion label generation.
    
    \item Processed results are saved to CSV and stored in the warehouse, later used for ML training and predictions.
\end{itemize}

\subsection{Modeling and Prediction}

\begin{itemize}
    \item Features: \texttt{v\_Vel}, \texttt{v\_Acc}, \texttt{Space\_Headway}, \texttt{Time\_Headway}
    \item Labels: Congestion level (generated via KMeans)
    \item Model: Random Forest with $n\_estimators = 100$
    \item Evaluation: Accuracy and Macro F1-score
\end{itemize}

\subsection{Performance and Scalability}

\begin{itemize}
    \item Total Records Processed: 11 million
    \item Peak Throughput: $\sim$320,000 records/min
    \item Average End-to-End Latency: 3.2 seconds per batch (100,000 records)
    \item Resource Usage: 65–75\% CPU, 8.2 GB of 12 GB RAM
    \item Stress Testing: Pushing all records at once caused a 10\% increase in processing time vs. chunked ingestion, due to Kafka-Spark consumer lag
\end{itemize}

\subsection{Key Contributions}

\begin{itemize}
    \item Demonstrated a scalable, Docker-based data pipeline without reliance on physical IoT sensors
    \item Introduced real-time congestion classification using unsupervised and supervised learning
    \item Validated system performance through stress testing and resource monitoring
    \item Delivered an integrated backend and frontend for live traffic insight visualization
\end{itemize}

\section{Data Visualization}

The following figures illustrate insights gained from early data processing and visualization steps:

\begin{figure}[htbp]
  \centering
  \includegraphics[width=0.48\textwidth]{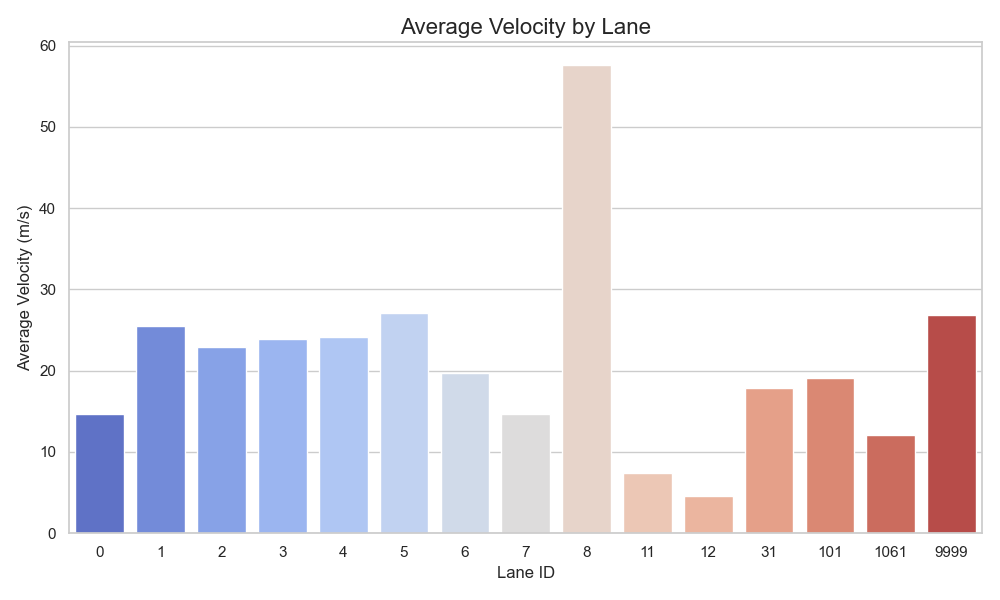}
  \includegraphics[width=0.48\textwidth]{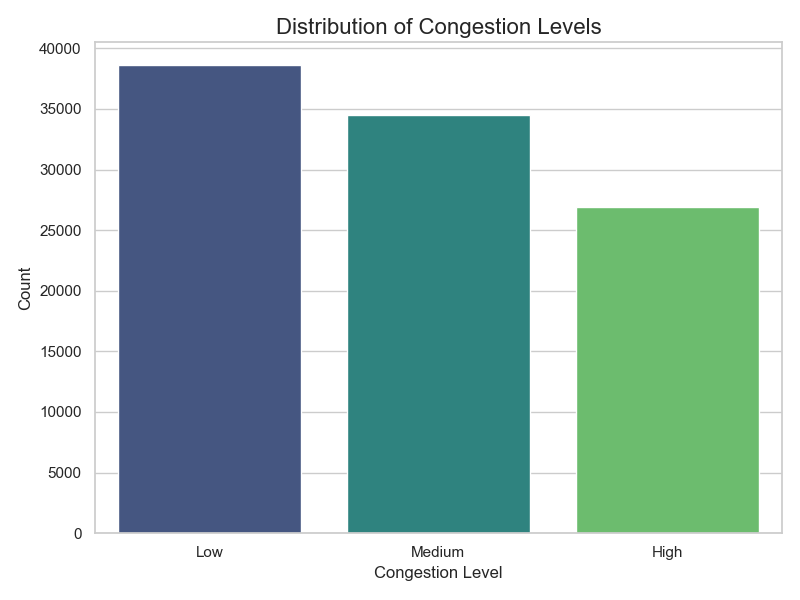}
  \caption{Left: Average velocity across lanes. Right: Distribution of congestion levels.}
  \label{fig:velocity-congestion}
\end{figure}

\begin{figure}[htbp]
  \centering
  \includegraphics[width=0.48\textwidth]{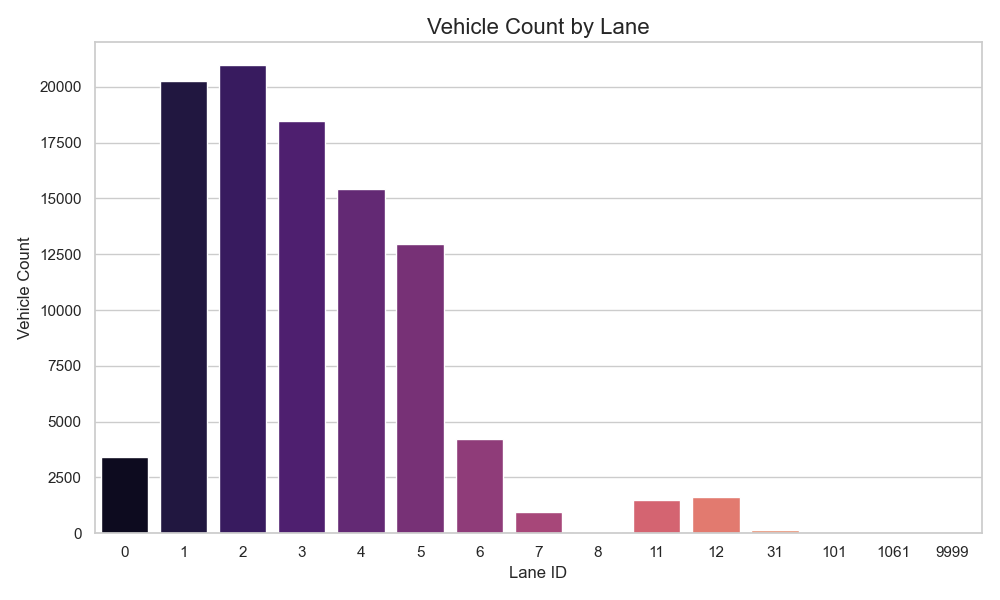}
  \includegraphics[width=0.48\textwidth]{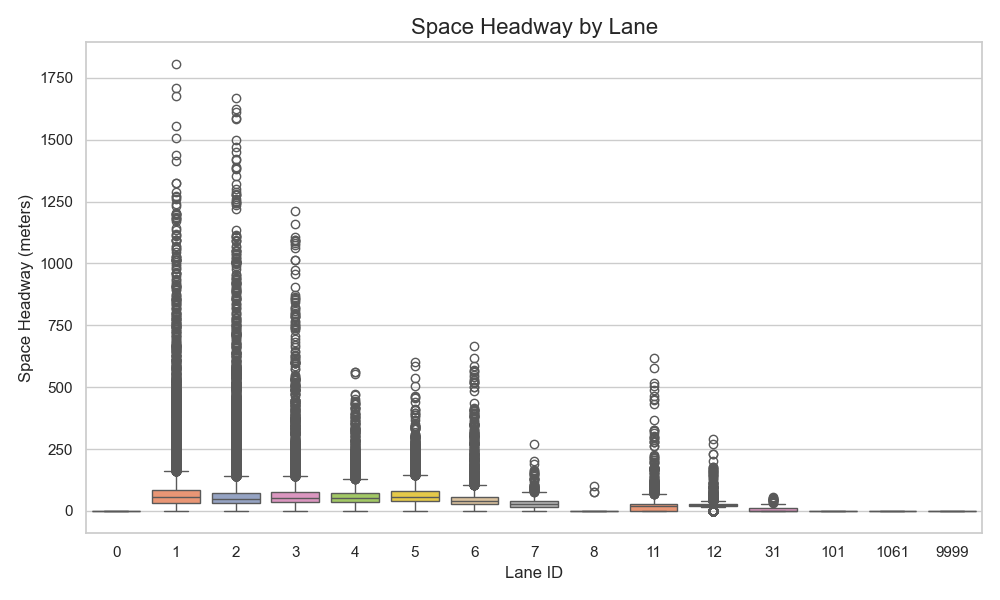}
  \caption{Left: Vehicle count per lane. Right: Average space headway by lane.}
  \label{fig:vehicle-space}
\end{figure}

\section{System Performance and Evaluation}

To assess the efficiency, scalability, and robustness of the \textit{CityPulse} pipeline, we conducted a series of performance evaluations under simulated high-load conditions. The system was tested with 11 million synthetic traffic records \cite{mousa2022deep} to measure throughput, latency, and resource usage during peak operations.

\subsection{Throughput and Latency}

The system demonstrated strong performance when processing large volumes of real-time data:
\begin{itemize}
    \item \textbf{Total Records Processed:} 11,000,000
    \item \textbf{Peak Throughput:} $\sim$320,000 records per minute
    \item \textbf{Average End-to-End Latency:} 3.2 seconds per batch (100,000 records)
\end{itemize}

These results reflect the pipeline’s ability to deliver near-real-time processing using Kafka for ingestion and Apache Spark Structured Streaming for transformation, supported by batch-wise optimization.

\subsection{System Resource Utilization}

Resource usage was monitored throughout the execution of the pipeline using Docker-level metrics:
\begin{itemize}
    \item \textbf{CPU Utilization:} 65\% to 75\% on average
    \item \textbf{Memory Consumption:} 8.2 GB (out of 12 GB available)
\end{itemize}

This indicates that the system remained stable under pressure, with sufficient headroom for scaling via container orchestration or cloud deployment.

\subsection{Stress Testing and Ingestion Strategies}

Two ingestion strategies were compared to evaluate their effect on system stability:
\begin{itemize}
    \item \textbf{Full Ingestion:} Injecting all 11M records in a single push caused a 10\% increase in processing time due to consumer backlog and disk I/O delays.
    \item \textbf{Chunked Ingestion:} Streaming data in batches of 500,000 records reduced memory pressure and improved throughput consistency.
\end{itemize}

These results show that chunked ingestion is more efficient for managing resource constraints and ensuring continuous flow in real-time systems.

\subsection{Identified Bottlenecks}

The evaluation revealed several points where performance degradation could occur:
\begin{itemize}
    \item \textbf{Kafka Consumer Lag:} Spark streaming occasionally fell behind Kafka during high-volume ingestion.
    \item \textbf{Spark Shuffle Overhead:} Memory usage peaked during transformations and clustering operations.
    \item \textbf{Temporary Storage I/O Delay:} Writing large batches to disk introduced minor latency before warehouse persistence.
\end{itemize}

Mitigation strategies included increasing Kafka partitions, tuning Spark batch intervals, and applying Snappy compression on Kafka messages to optimize I/O throughput.

\subsection{Evaluation Summary}

Overall, the system handled large-scale data ingestion and processing reliably and efficiently. With its modular design, Docker-based deployment, and ability to maintain stable latency and resource usage under stress, \textit{CityPulse} demonstrates practical scalability for real-world traffic analytics applications, particularly in environments with limited infrastructure.

\section{Machine Learning Evaluation}

The congestion classification module in \textit{CityPulse} is built using a Random Forest classifier \cite{zhang2020urban}, trained on key engineered features: \texttt{v\_Vel}, \texttt{v\_Acc}, \texttt{Time\_Headway}, and \texttt{Space\_Headway}. Congestion labels (High, Medium, Low) were initially generated through unsupervised KMeans clustering, and used as training targets in the supervised model.

\subsection{Feature Importance}

\begin{figure}[htbp]
    \centering
    \includegraphics[width=0.5\textwidth]{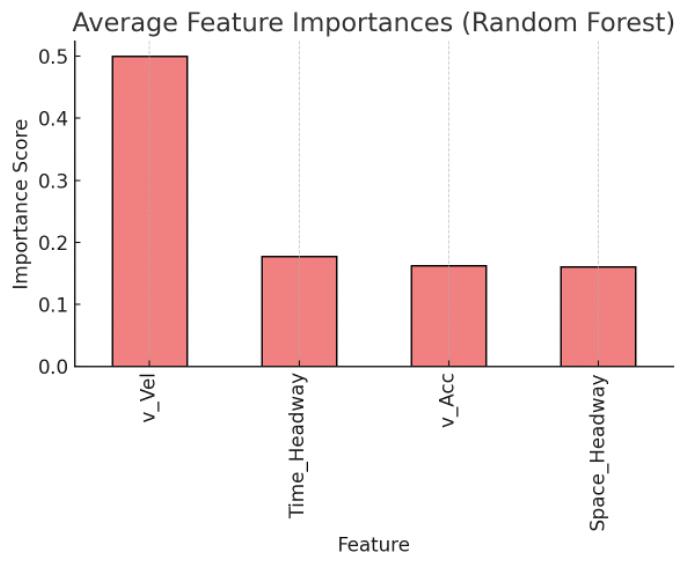}
    \caption{Average Feature Importances from Random Forest Classifier}
    \label{fig:feature-importance}
\end{figure}

Figure~\ref{fig:feature-importance} illustrates the average feature importances computed by the trained Random Forest model. The most influential variable was \texttt{v\_Vel}, contributing over 50\% to the model's predictions, followed by \texttt{Time\_Headway}, \texttt{v\_Acc}, and \texttt{Space\_Headway}. This confirms the importance of velocity as a leading indicator of traffic congestion.

\subsection{Classification Metrics}

\begin{figure}[htbp]
    \centering
    \includegraphics[width=0.5\textwidth]{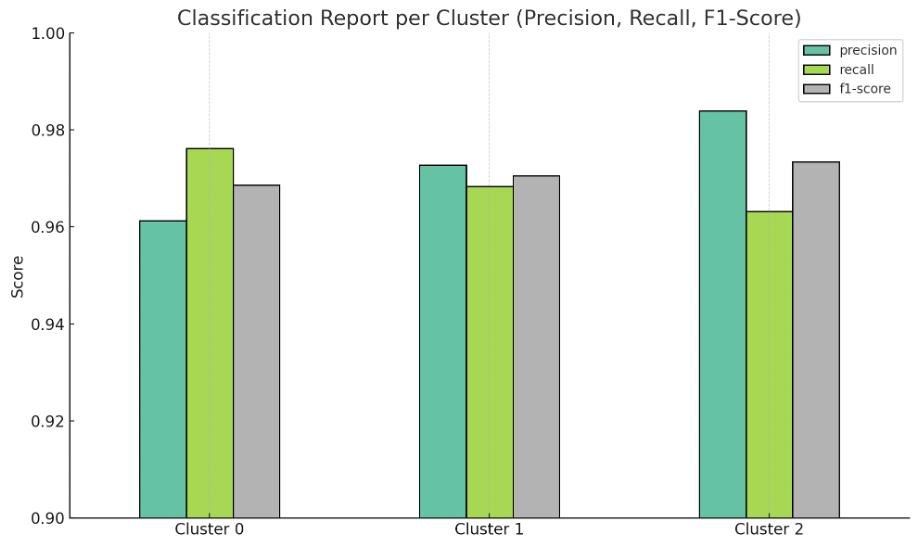}
    \caption{Classification Report Per Cluster (Precision, Recall, F1-Score)}
    \label{fig:cluster-metrics}
\end{figure}

Figure~\ref{fig:cluster-metrics} presents the precision, recall, and F1-score per cluster. The model achieved macro-averaged scores above 0.96 across all three congestion classes, indicating balanced performance and minimal bias between class labels. 

\subsection{Stability Across Batches}

\begin{figure}[htbp]
    \centering
    \includegraphics[width=0.5\textwidth]{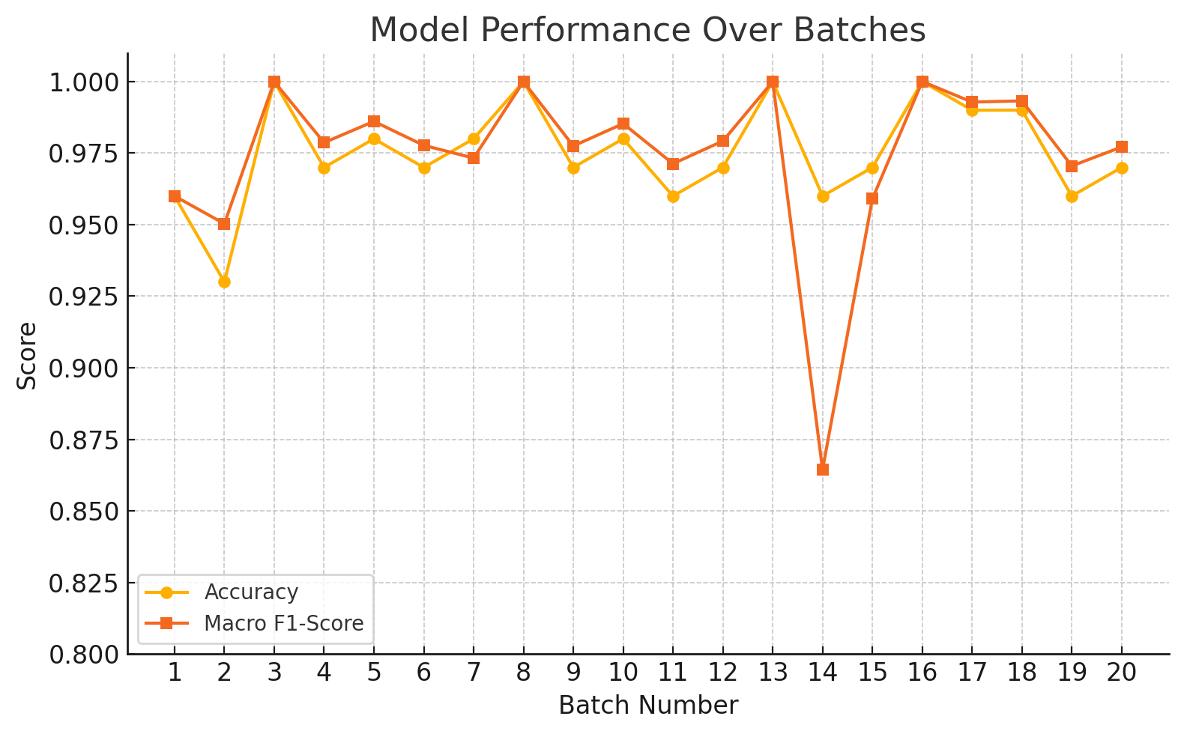}
    \caption{Model Accuracy and F1-Score Over Sequential Batches}
    \label{fig:batch-performance}
\end{figure}

To ensure robustness, the model was evaluated over 20 sequential data batches. As shown in Figure~\ref{fig:batch-performance}, the accuracy and macro F1-score consistently remained above 0.95, with only a single temporary drop due to noisy input in batch 14. The model quickly recovered, demonstrating resilience to anomalies.

\subsection{Confusion Matrix}

\begin{figure}[htbp]
    \centering
    \includegraphics[width=0.5\textwidth]{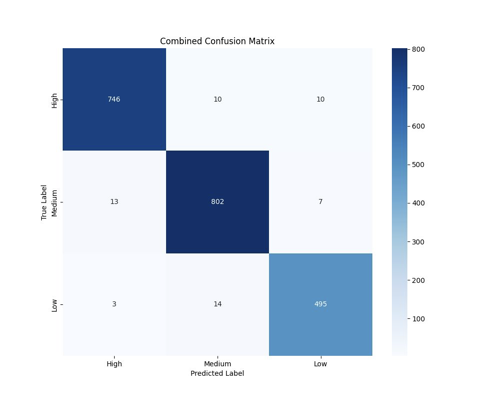}
    \caption{Combined Confusion Matrix Across All Batches}
    \label{fig:confusion-matrix}
\end{figure}

The confusion matrix in Figure~\ref{fig:confusion-matrix} shows high agreement between predicted and true congestion labels, especially for the "Medium" class. Misclassifications were minimal and primarily occurred between adjacent congestion levels (e.g., High vs. Medium), which is expected in real-world traffic transitions.

\subsection{Summary}

The evaluation confirms that the Random Forest model performs reliably with high precision and recall, is robust to batch-level variations, and leverages meaningful features for congestion prediction \cite{sun2018parallel}. These results demonstrate the feasibility of using simulated data for training predictive traffic models in the absence of physical sensors.

\section{Conclusion}

This work presented \textit{CityPulse}, a scalable, real-time traffic congestion analytics pipeline designed for smart city applications in data-scarce environments. By simulating 11 million traffic records and processing them through a fully containerized pipeline using Kafka, Spark, and Docker, we demonstrated how modern open-source technologies can replace traditional sensor-based infrastructure.

The system achieved high throughput and low latency under stress conditions, and the modular architecture ensured fault-tolerant, reproducible deployments. The temporary storage layer and batch-wise Kafka ingestion helped mitigate performance bottlenecks. Our machine learning evaluation, powered by Random Forests trained on features such as velocity and headway metrics, achieved macro F1-scores above 0.95 across all congestion classes.

\textit{CityPulse} shows that even without physical sensors or cloud infrastructure, meaningful real-time traffic insights can be generated with lightweight, locally deployable tools. This opens up opportunities for municipalities in developing regions to leapfrog traditional infrastructure and implement data-driven urban mobility planning.

\section{Future Work}

While \textit{CityPulse} provides a robust foundation for real-time traffic analytics using simulated data, several avenues remain for future development and deployment:

\begin{itemize}
    \item \textbf{Integration with Real-Time Sensors:} Future iterations of the system will incorporate real sensor streams such as GPS data from mobile devices, loop detectors, or roadside cameras to replace or augment synthetic data.
    
    \item \textbf{Edge and Cloud Deployment:} The current Dockerized setup can be extended to edge devices (e.g., Raspberry Pi, Jetson Nano) for localized traffic analysis, or scaled on cloud platforms (e.g., AWS EC2, Kubernetes) to handle larger geographic regions.
    
    \item \textbf{Map-Based Visualization:} A future frontend enhancement will include map-based traffic visualization with dynamic congestion overlays and time-based playback to improve usability for city planners and emergency responders.
    
    \item \textbf{Multi-Class and Temporal Forecasting:} Future modeling work will explore time-series models and deep learning techniques (e.g., LSTM, TCNs) to forecast traffic congestion levels minutes or hours ahead, not just classify current status.
    
    \item \textbf{Policy Simulation and Recommendation:} The system could be extended to test the effects of different traffic policies (e.g., signal timing, road closures) using simulation-based inference or reinforcement learning.
    
    \item \textbf{Localization for Developing Regions:} Future deployments will emphasize regional tuning (e.g., Cameroon's urban topology) and language localization to ensure accessibility and adoption by local agencies and communities.
\end{itemize}

Overall, \textit{CityPulse} has laid the groundwork for accessible, intelligent traffic management systems and can evolve into a critical decision-support tool for urban mobility planning in low-resource settings.

\bibliographystyle{alphaurl}
\bibliography{references}

\appendix

\end{document}